\documentclass{aa}
\usepackage{times}
\usepackage{epsfig}

\thesaurus{2 
	   (12.03.3;  
	    12.12.1)} 

\title{Clustering of galaxies at faint magnitudes}

\author{J.U. Fynbo \inst{1,2}
   \and W. Freudling \inst{2,3}
   \and P. M\o ller \inst{2} 
   }

\institute{
           Institute of Physics and Astronomy,
           \AA rhus University, DK-8000 \AA rhus C., Denmark
           \and
           European Southern Observatory,  
           Karl-Schwarzschild-Stra\ss e 2,
	   D-85748, Garching bei M\"unchen, Germany
	   \and
	   Space Telescope -- European Coordinating Facility,
	   Karl-Schwarzschild-Stra\ss e 2,
	   D-85748, Garching bei M\"unchen, Germany
           }

\offprints{J.U. Fynbo}
\mail{jfynbo@ifa.au.dk}

\date{Received 28 May 1999 / Accepted 21 December 1999}

\begin{document}

\maketitle

\begin{abstract}

Significant uncertainties exist in the measured amplitude of the
angular two-point correlation function of galaxies at magnitudes
$I\approx26$ and fainter.  Published results from HST and ground-based
galaxy catalogs seem to differ by as much as a factor of 3, and it is not
clear whether the correlation amplitude as a function of magnitude
increases or decreases in the faintest magnitude bins. In order to
clarify the situation, we present new results from both ground-based
and HST galaxy catalogs.  The angular two-point correlation function
as a function of limiting R and I magnitudes was computed from a
galaxy catalog created from the Hubble Deep Field - South (HDF-S)
WFPC2 image.  The measured amplitudes of the correlation at an angular
separation of 1 arcsec are consistent with those measured in the
Northern counter part of the field. The flanking fields (FF fields) of
the Hubble deep fields were used to extend the magnitude range for
which we compute correlation amplitudes towards brighter magnitude
bins.  This allows easier comparison of the amplitudes to ground based
data. The ESO NTT Deep Field catalog was used to measure the
correlation at similar magnitudes from ground based data. We find that
the measured correlation from both the HST and ground based samples
are consistent with a continuously decreasing clustering amplitude
down to the faintest magnitude limits.  Finally, the newly measured
correlation amplitudes as a function of magnitude limit were compared
to previously published measurements at larger separations. For this
comparison, the correlation function was approximated by a power law
with an index of 0.8. The scatter in the correlation amplitudes is
too large to be explained by random errors. We argue that the most
likely cause is the assumption that the shape of the correlation
function does not depend on the magnitude limit.

\end{abstract}

\section{Introduction}

The angular correlation function of galaxies $w(\theta)$ is still the 
most commonly used tool for investigating the evolution of clustering
at high redshift . While a more direct measurement of the two point
correlation is significantly more powerful (see e.g. Le F\`evre et
al. 1996), its computation requires deep magnitude limited redshift
surveys, which currently are not feasable at the faintest reachable 
magnitude limits. By contrast,
the angular correlation function can easily be computed from
photometric galaxy catalogs alone. With the availability of the Northern
Hubble Deep Fields (HDF-N, Williams et al. 1996), the angular
correlation function has been computed to an R magnitude of 29 (Villumsen 
et al. 1996, hereafter VFC, Colley et al. 1996). 
These studies suggest that the amplitude of the
correlation function continuously decreases with the magnitude limit of
the sample over a magnitude range of more than 10 magnitudes down to
the faintest magnitude limits probed so far. The amplitude at an
angular separation of one arcsec, $w(1'')$, seems to follow closely
a power law $w(1'')\propto10^{-0.27R}$(e.g.  Brainerd et al. 1995).

Despite the apparent simplicity of computing the angular correlation
function, a significant controversy has arisen about the magnitude
limits at which the correlation function flattens. Measuring the
correlation amplitude at a separation of 1 arcmin, Brainerd \& Smail
(1998) found that the correlation amplitude reaches a minimum for
samples with limiting I magnitude of about 23 and stays flat for even
deeper magnitude limited samples. On the other hand, the deeper
HDF-N measurements at a separation of 1 arcsec found a continuously
decreasing clustering amplitude down to the faintest magnitude
limits.  This discrepancy could be due to a number of different
reasons such as biased field selection, different redshift selection
through differences in the bandpasses, systematic errors in the
determination of the clustering amplitude (e.g. due to gradients in
the sensitivity of the detector), or simply random fluctuations in the
correlation amplitudes. However, it could also mean a true
discontinuity of the shape and/or amplitude of the correlation
function at a magnitude of $I\approx26$ or $R\approx27$.

Before suggesting the existence of a discontinuity of the correlation 
amplitudes, several tests of the results should be carried out. First of 
all, both the ground based results and
the HDF-N results should be verified by independent
samples. Secondly, any apparent discontinuity should be tested with a
single sample which covers the relevant magnitude range. Up to now,
available ground based have not been not deep enough to overlap with the HDF
results, which are only available for magnitudes fainter than
$R\approx26$. The current work addresses both of these points. We have
used the southern HDF field to obtain an independent verification of the
previous HDF-N results. In addition, we have used the ``flanking
fields'' of the Northern and Southern HDFs  (Williams et al. 1998) to derive a
catalog over a larger area and thus allowing us to compute the
$w(\theta)$ at brighter magnitudes from an HDF-like sample. Finally, we
have used a sample of galaxies detected in a deep ground based
image, the so-called ESO NTT deep field (Arnouts et al. 1999) to compute
$w(\theta)$ from ground-based data at faint magnitudes. The combined
data sets allow us to derive correlation amplitudes from both HST
and ground based data at identical magnitude bins, and thereby search
for discontinuities in the correlation amplitudes as a function of magnitude.

In Sect.~\ref{HDF}, we present the results from the HDF fields
which include the Northern HDF, the Southern HDF and the flanking fields. 
In Sect.~\ref{nttfield}, we present new results from the ESO NTT deep
field. In Sect.~\ref{comparison}, we investigate the correlation
amplitude as a function of magnitude by combining the new data with previous
estimates in the literature. Finally, in Sect.~\ref{conclusion}, we 
discuss the results and present our conclusions.

\section{HDF samples}\label{HDF}

\subsection{HDF South : The galaxy catalog}
\label{catalog}

\begin{figure}
   \epsfig{file=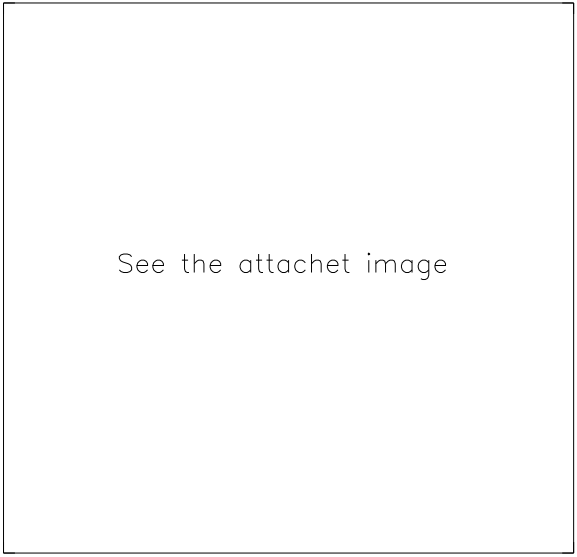, width=8cm}
   \caption{The WFPC2 HDF-S field showing the the used areas of each
            of the three Wide Field camera CCDs.}
   \label{used_area}
\end{figure}

We used the ``drizzled'' HDF-S images distributed by the 
{\it Space Telescope -- European Coordinating Facility} archive. The
galaxy catalog was generated using the SExtractor program (Bertin
\& Arnouts 1996). The catalog of Clements \& Couch (1996) was used 
by VFC to compute $w(\theta)$ from the HDF-N images. In order to be
able to compare our results with the results obtained from the HDF-N
we use the exact same extraction parameters as input to SExtractor as
those described in Clements \& Couch. We use a minimum object
extraction area of 30 pixels and a detection threshold of 1.3$\sigma$
above the background.  As a detection image we use a master frame
calculated as the sum of the combined images taken with the F606W and
F814W filters, which are similar to the R and I passbands
respectively. As a detection filter we use a top-hat filter of 30
connected pixels. We checked the reliability of the extraction
parameters by comparing the master frame with the output object image,
which SExtractor optionally provides. We could not detect any tendency
to find spurious objects in the vicinity of bright objects. The useful
part of the combined mosaic frames are different between the HDF-N and
HDF-S because the WFPC2 fields of view are slightly rotated relative
to the mosaiced image which is aligned with the north/south
direction. Therefore, we re-defined for each of the WF2, WF3 and WF4
CCDs useful areas from visual inspection of the master
frame. Fig.\ref{used_area} shows the chosen areas. Only objects
detected
within the shown boundaries were used in the analysis. In order
to excluded stars from the galaxy catalog we set an upper limit to the
neural network star parameter in the output SExtractor catalog of
0.98, which excluded 20 objects. This is consistent with the number of 
stars expected for a field with the galactic latitude and depth as the
HDF-S.  Also we exclude saturated stars found by visual inspection
from the galaxy catalog. There are not very bright stars in the HDF-S
so we did not mask out any regions within the area shown in 
Fig.~\ref{used_area}. We chose to use the Automatic Aperture
Magnitudes (AAM) instead of the Corrected Isophotal Magnitudes (CIM),
since the latter proved unstable. As well AAM as CIM are intended to
give an estimate of the total flux of an object. The AAM is measured
using an elliptical aperture with minor axis $b = 2.5 r_1 \epsilon$
and major axis $a = 2.5 {r_1 \over
\epsilon}$, where $r_1$ is the first moment of the light distribution
and $\epsilon$ is the ellipticity.

The resulting number of galaxies in the galaxy catalog is 1346 with 456 
in the WF2 field, 530 in the WF3 field and 360 in the WF4 field. 

\begin{figure}
   \epsfig{file=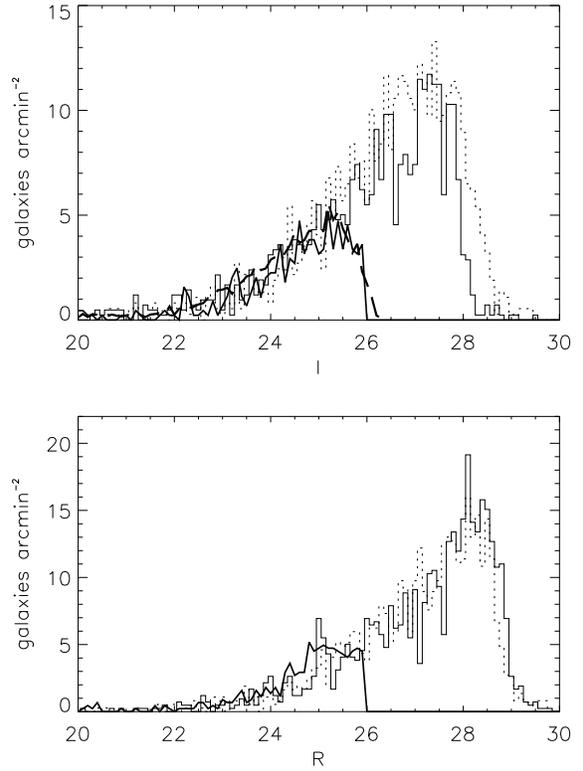, width=8cm}
   \caption{Comparison of the galaxy counts in 0.1mag bins in the HDF-N
and HDF-S fields in the I and R bands. The solid 
histogram shows the counts for HDF-N and the dashed histogram shows the 
counts in HDF-S. In the top plot the fat long dashed line shows the 
galaxy counts in the HDF flanking fields and the fat solid line the 
counts in NTT SUSI Deep Field. The fat solid line in the lower plot shows 
the galaxy counts for the NTT SUSI Deep Field.}
   \label{counts}
\end{figure}

As zero-points for the photometry we use those released by the STScI
with the WFPC2 HDF-S data for the VEGAMAG system. In the following we
shall refer to the F606W and F814W as R and I respectively.
Fig.~\ref{counts} compares the galaxy counts for HDF-N and HDF-S.  It
is seen that the distributions are similar. In particular,
the magnitude bins at which the number counts drop significantly  
due to incompleteness in the two samples are almost identical. The 
basic properties of the galaxy catalogs from the two fields as well 
as from other fields to be discussed below, are listed in Table 1.

\begin{table}
 \begin{center}
\caption{Survey data for the deep fields used in the determination of 
the twopoint correlation function}
 \begin{tabular}{@{}lcccccc}
  field & Area         & I(complete) & R(complete) \\
  \hline
        & arcsec$^{2}$ &             &             \\
  \hline
HDF-S    &  3.7        &   28.0      &  28.5       \\
HDF-S FF   & 33.2        &   25.5      &   -         \\
HDF-N    &  4.2        &   28.0      &  28.5       \\
HDF-N FF   &  8.7        &   25.5      &   -         \\
NTT     &  4.4        &   25.5      &  25.5       \\
\hline
\end{tabular}
\label{fields}
\end{center}
\end{table}

\subsection{Correlation function from the HDF south catalog}
\label{w}
   We extract from the galaxy catalog eight R-magnitude-limited 
samples with limiting magnitudes from R=25.5 to R=29.0 in 0.5 mag steps. 
As for the HDF-N, galaxies brighter than R=23 are excluded.  
 
   The angular correlation function $w(\theta)$ is estimated using the 
optimal estimator described by Landy \& Szalay (1993) :

\begin{equation}
w(\theta) = \frac{DD - 2DR + RR}{RR},
\end{equation}

\noindent
where DD is the number of galaxy-galaxy pairs, DR the number of
galaxy-random pairs and RR the number of random-random pairs at the
separation $\theta$. We generate a set of 32000 random points 
distributed according to the Poisson distribution. The number of
galaxy-random pairs and random-random pairs at a given magnitude 
limit are normalized to the total number of galaxy-random and 
random-random pairs respectively. The errors are calculated from
Poisson statistics.

   Since the mean density of galaxies has to be estimated from the
sample itself, the integral of the correlation function over the survey
area is forced to zero. This reduces the correlation $w(\theta)$ function 
by an amount $C$ given by the integral :

\begin{equation}
C = \frac{1}{\Omega^2} \int\!\!\!\int d\Omega_1\, d\Omega_2\: w(\theta),
\end{equation}

\noindent 
where $\Omega$ is the solid angle of the survey area. $C$ is 
commonly referred to as the 'integral constraint`. Assuming that 
$w(\theta)$ is a power law,

\begin{equation}
w(\theta) = A\theta^{-\gamma+1},\:\:\:\:\:\gamma=1.8\:,
\end{equation}

\noindent 
then $C = 0.078(0.077,0.080)A$ for the used parts of the WF2(WF3,WF4) 
area shown in Fig.\ref{used_area}. 

\begin{figure*}
\epsfig{file=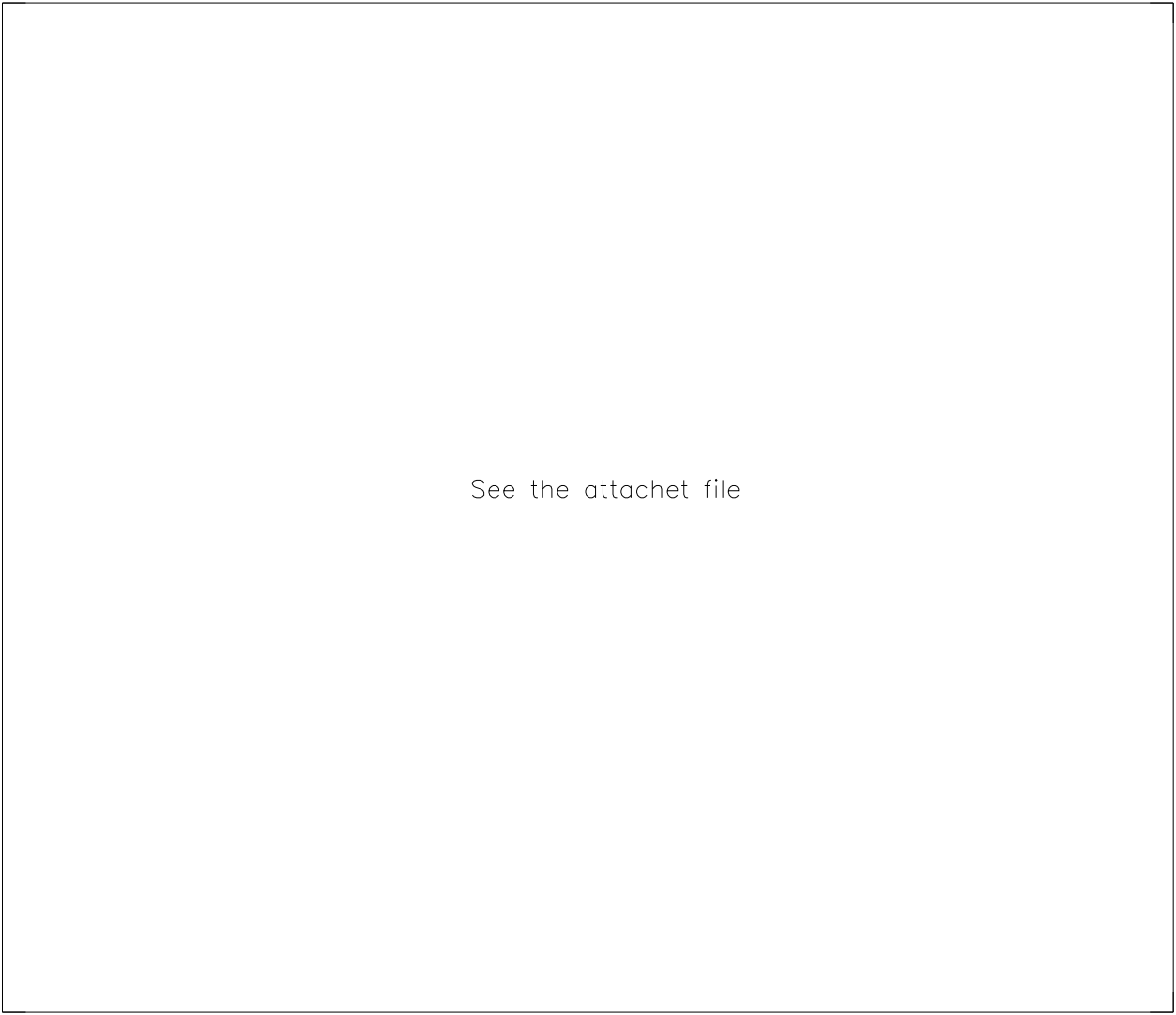,width=17cm,height=15cm}
\caption{
The measured angular correlation function for the eight magnitude bins.
The limiting magnitude for each bin has been indicated on each plot.
Also shown is the best fit to a function of the form given in Eq. 
(4) and the plus and minus 1$\sigma$ fits. 
}
\label{results}
\end{figure*}

  Fig.\ref{results} shows the observed $w(\theta)$ for the eight 
magnitude limits. The error bars represent 1 $\sigma$ Poisson errors.
The amplitude of $w(\theta)$ is determined by fitting

\begin{equation}
w(\theta) = A\theta^{-\gamma+1}\:-\,C,\:\:\:\:\:\gamma=1.8\:,
\end{equation}

\noindent 
which takes into account the integral constraint. 

\begin{figure}
\epsfig{file=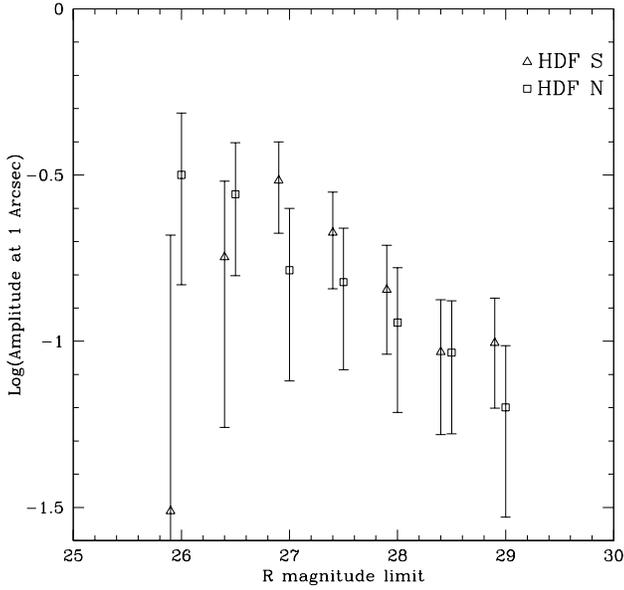,width=8.5cm}
\caption{Comparison for the R band of the amplitude of the angular 
correlation function in seven magnitude bins for HDF-N (triangles) and 
HDF-S (squares). The data points for the HDF-S have been shifted 0.1 mag
to the left to allow an easier comparison. In all bins do the 
measurements agree within 1$\sigma$.
}
\label{compare}
\end{figure}

\begin{figure}
\epsfig{file=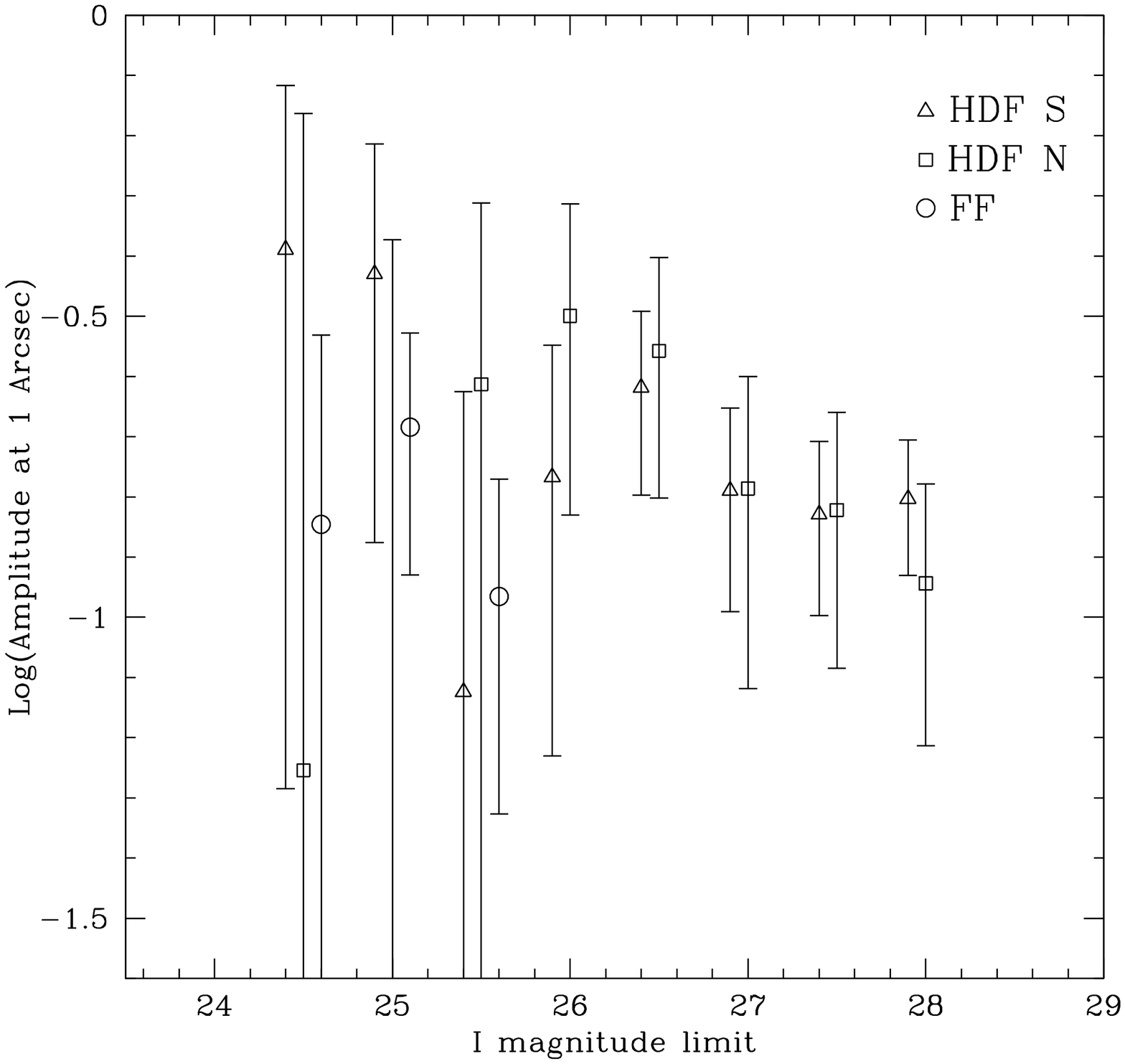,width=8.5cm}
\caption{Comparison for the I band of the amplitude of the angular 
correlation function
for seven magnitude bins for HDF-N (triangles), HDF-S
(squares) and for the three first bins also the flanking fields (FF, 
circles). The data points for the HDF-S have been shifted 0.1 mag to 
the left and the points for the flanking fields 0.1 mag to the right to 
allow an easier comparison. In all bins do the measurements agree within 
1$\sigma$.
}
\label{compareI}
\end{figure}

\subsection{Flanking fields}

The high resolution deep images of HDF fields are uniquely suited 
to investigate the correlation function at the faintest
magnitudes. However, their small field of view includes too few brighter
galaxies to provide useful overlap with ground based measurements 
of the correlation function. This situation can be improved by 
taking advantage of the HDF ``flanking fields''. These fields are 
contiguous to the HDF fields proper and have been taken during the HDF 
observing campaigns. We use the two deepest of the HDF-N flanking fields 
and the nine HDF-S flanking fields. 

The HDF proper fields were chosen to avoid bright galaxies. It could
be argued that this choice biases the measured correlation amplitude
(Brainerd \& Smail 1998). As the flanking fields have been chosen with 
proximity to the HDF fields as the only criterion, no bias related to 
the HDF proper field is present.
 
  The flanking fields were only imaged with the F814W filter. There
is a total of 14 fields. Three of the southern fields contained very 
bright stars. Rather than masking out these stars we decided not to use 
these frames in the analysis. Catalogs from the remaining 11 fields 
were generated in the
same manner as for the HDF-S field. The average number counts as a
function of magnitude are shown in the upper plot in
Fig.~\ref{counts}. For magnitudes up to an I magnitude of about 25
the number per area is consistent with the one in the deep fields. The
larger area therefore increases the sample in the brighter magnitude
bins by almost an order of magnitude. The relevant parameters
are again summarized in Table 1. 

Finally, we measure the two point correlation function in the three
bins $23<$I$<24.5$, $23<$I$<25.0$, and $23<$I$<25.5$ in the same way as
described in Sect.~\ref{w}.

\subsection{The correlation function from the combined Hubble Deep Fields}

In Fig.\ref{compare} we compare the HDF-S R band results with the results 
derived
by VFC for the HDF-N. The triangles with 1$\sigma$ error bars show
the results for the HDF-S and the squares with 1$\sigma$ error
bars show the results from the HDF-N. For all magnitude bins the
measurements from the two fields agree within 1$\sigma$. We
conclude that any differences are due to random fluctuations, and
therefore combine all the available data from the HDF-N and HDF-S
projects to obtain our final estimate of the correlation
amplitudes from the HDF fields. In Fig.\ref{compareI} we perform
the same comparison for the I band, this time also including the
measurements from the flanking fields. The measurements are in
all bins consistent within 1$\sigma$, and we again combine all the
available data from the HDF-N and HDF-S projects including the
Flanking fields to obtain our final estimate of the I band correlation
amplitudes from the HDF fields.

For the R band we determine $w(\theta)$ in the 8 magnitude bins
R$<$25.5, R$<$26 ..., R$<$29. In  these bins the number counts are
not significantly influenced by incompleteness. For the I band we
determine $w(\theta)$ in the 8 magnitude bins I$<$24.5,
I$<$25,...,I$<$28. The flanking fields are here included in the
measurement of $w(\theta)$ in the first three bins. 

The determination of $w(\theta)$ and the power law fit is done 
as described in Sect.~\ref{w}. We detect a positive correlation 
amplitude in all 8 bins in both R and I.

\section{The ESO-NTT deep field}\label{nttfield}

In order to verify that the correlation amplitudes we measured in the
HST WFPC2 frames are not in some way due to properties of the
WFPC2 instrument or HDF observing or reduction procedure, it is
desirable to compare to a catalog of galaxies derived from deep
high-resolution ground based observations. The ESO-NTT field (Arnouts
et al. 1999), for which the images are publicly available, is
suitable for this purpose. A catalog of galaxies for this field
has been provided by S. D'Odorico at the ESO website at
http://www.eso.org/ndf. The field provides deep multi-colour imaging
taken with good sampling of the PSF. The depth of the images is about
26.7 in R and 26.3 in I. The effective seeings of the co-added images
are about 0.7-0.8 arcsec.

We determine the amplitude of $w(\theta)$ in the magnitude intervals
$22$--$24.5$, $22$--$25.0$ and $22$--$25.5$ for both the I and R
filters. We use an area of 4.4 arcmin$^2$ confined by the pixel values
$40<$x,y$<1020$ (chosen to avoid edge-effects). For this area we
derive an integral constraint of $C=0.0468A$.  Due to the small size
of the NTT SUSI Deep Field we cannot determine the amplitude of
$w(\theta)$ very precisely.  Nevertheless, we do detect a positive
correlation signal in all but one of the magnitude bins. The measured
amplitudes of $w(\theta)$ are in all bins consistent with the
measurements in the two HDFs within 1$\sigma$ (see
Fig.~\ref{compare_ground} and Fig.~\ref{compare_groundI}).

Stellar contamination of the sample can reduce the measured
correlation amplitude. For a sample with a fraction $f_{\rm star}$ of
galaxies, the measured correlation amplitude is reduced by a factor
$\propto (1- f_{\rm star})^{-2}$. Fortunately, the excellent seeing
allows a firm galaxy/star discrimination. We follow Arnouts et al. in
classifying as stars all objects with SExtractor classifier larger
than 0.9 in the I band. This procedure removes about 5\% of the
objects in the field. For objects fainter than 24 the SExtractor
classifier is not efficient, so we assume conservatively that about
5\% of the objects used to measure the correlation amplitude are
stars. In fact, the star counts are shallower at faint magnitudes than
that of galaxies (Reid et al. 1996, their Fig. 1), which means that
the star fraction at $I>24$ must be smaller than for
$I<24$. Therefore, a conservative upper limit of the error introduced
by stellar contamination of the catalog is 10\%, which is
significantly smaller than the random uncertainties.

\section{The correlation amplitude as a function of magnitudes}\label{comparison}

The correlation 
amplitudes presented above overlap neither in magnitude nor scale with
most such measurements from ground based data. Investigations of the
correlation amplitude as a function of magnitudes which use these data
therefore require knowledge of the power law index $\delta$ in order
to compare amplitudes measured at different separations. The HST data
presented here are consistent with the usual $\delta=0.8$ power law for
angular separations between 1 arcsec and 1 arcmin, but do not rule
out steeper of flatter slopes. On the other hand, ground based data
have been used to derive the shape of the correlations function at
separations between 0.5 and 5 arcminutes (e.g. Postman et al. 1998),
but it is not clear whether the slopes can be extrapolated to smaller
separations. In the following investigation, we have adopted a slope of
0.8 in most cases.  The uncertainty in the slope is probably the
largest source of uncertainty in the comparison of measurements from
different surveys.

Fig.~\ref{compare_ground} shows the measured amplitudes of $w(\theta)$
at 1 arcsec as a function of the limiting magnitudes of surveys in the
R band from measurements by Brainerd et al. (1995), Roche et al.
(1993), Couch et al. (1993) and Woods \& Fahlman (1997).  Where
necessary (data from Roche et al. and Couch et al.), we have scaled
original correlation amplitudes quoted in each paper to the amplitude
at 1 arcsec by assuming a power law for the correlation function with
$\delta=0.8$. Similarly, Fig.~\ref{compare_groundI} shows the measured
amplitudes of $w(\theta)$ at 1 arcsec as a function of the limiting
magnitudes of the surveys in the I band. The measurements of
Postman et al. (1998), Brainerd \& Smail (1998), Woods \& Fahlman
(1997), Benoist et al.  (1999), Benoist et al. (in prep.) and
Lidman \& Peterson (1996) are included. Here, all the amplitudes
except those from Woods \& Fahlman (1997) and Postman et al. (1998) were
extrapolated to a separation of 1 arcsec with $\delta=0.8$. For
the Postman et al. (1998) measurements, the data at 1 arcmin and 0.5
arcmin were extrapolated using the slope measured by the authors in
the range 0.5 arcmin -- 5 arcmin.

Both in Fig.~\ref{compare_ground} 
and Fig.~\ref{compare_groundI}, we show theoretical predictions of the 
expected correlation amplitudes. The models are identical to those
shown by VFC, but with the assumed redshift distribution in the faint
magnitude bins derived from the photometric redshift distribution of
HDF-N galaxy by Fern\'andez-Soto et al. (1999). The median redshifts
adopted at magnitudes brighter than 26 in the R band are the same as 
those used by VFC. The models are specified by a present clustering 
length  $r_0$ and and evolution parameter $\epsilon$ (see VFC for details). 
The models shown have a present clustering length of $r_0=4.0 h^{-1}$ 
Mpc and assume no evolution of clustering (dotted line, $\epsilon$=0),
linear evolution of clustering (solid line, $\epsilon$=0.8) or 
non-linear evolution of clustering (dashed line, $\epsilon$=1.6). 

Several data sets shown in Fig.~\ref{compare_ground} and
Fig.~\ref{compare_groundI} are in conflict with each other at 
the 5-7$\sigma$ level. This reflects the fact that the main source of
errors are systematic errors, while the error bars in most cases only
include statistical errors. In particular at faint magnitudes,
systematic uncertainties are hard to quantify. The disagreements appear
to be larger in the I-band. This might be due to the fact that in the
I band, variations in the effective filter bandpasses result in
relatively large differences in the redshift distribution of extracted
galaxy samples. Also, the limiting magnitude on the abscissa is somewhat
depending on the limiting magnitudes at the {\it bright} end. However, as 
the galaxy counts ($N$) increases steeply with the limiting magnitude 
at the faint end (mag$_{lim}$, $\log N \propto 0.6$ mag$_{lim}$), the 
uncertainty due to different limiting magnitudes is small (the 
difference in the median magnitude of two galaxy samples with limiting 
magnitude 25 at the faint end and limiting magnitudes at the bright 
end of 20 and 24 is only about 0.5mag). More likely, however, the 
disagreement is related to
the extrapolation from 1 arcmin scales to 1 arcsec scales. In Postman et
al. (1998, their Table 3) it can be seen that the power $\delta$ seem to
decline at $I>21$ to smaller values than the canonical value 0.8
($\delta=0.7$ at $I=22$ and $\delta=0.5$ at $I=23$), 
especially at the smallest scales 0.5 arcsec $< \theta <$ 5 arcmin. 
A similar trend was noted by Benoist et al. (1999). Brainerd \& Smail  
(1998) find at best fitting value of $\delta=0.7$ at $I>24.5$.
A change in slope from $\delta=0.8$ to $\delta=0.5$ amounts to
factor of 3.4 difference in the extrapolation from 1 arcmin scales
to 1 arcsec scales. Due to these uncertainties, our current knowledge 
of the amplitude and shape of 
the correlation function in the range 1 arcmin $< \theta <$ 1 arcsec
at the faintest magnitude limits reachable from the ground, I=23--25,
is unfortunately still poor.

\section{Summary and conclusions}\label{conclusion}

{}From the data presented above we can draw the following
conclusions.

\begin{figure}
\epsfig{file=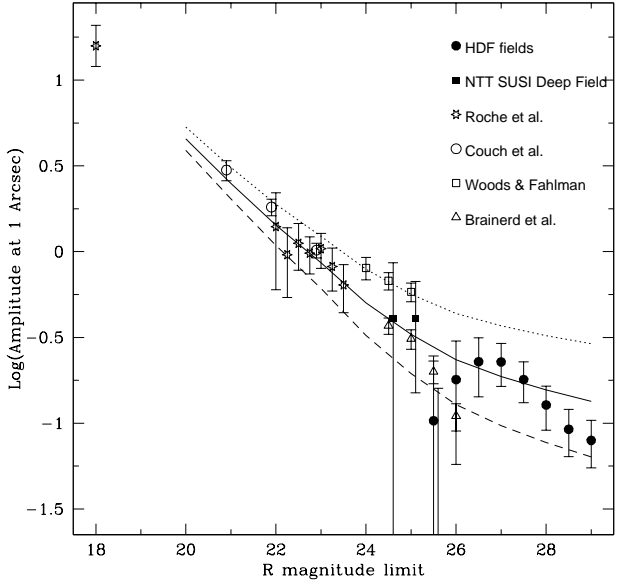,width=8.5cm}
\caption{The amplitude of $w(\theta)$ at 1 arcsec as a function of
the limiting R band magnitude for a number of ground based surveys and 
the measurements from the Hubble Deep fields presented in this paper.
The superimposed curves are theoretical predictions for a 
population of galaxies
with a present clustering length of $r_0=4.0 h^{-1}$ Mpc and assuming
no evolution of clustering (dotted line), linear evolution (solid line)
as well as non-linear evolution of clustering (dashed line).  }
\label{compare_ground}
\end{figure}

\begin{figure}
\epsfig{file=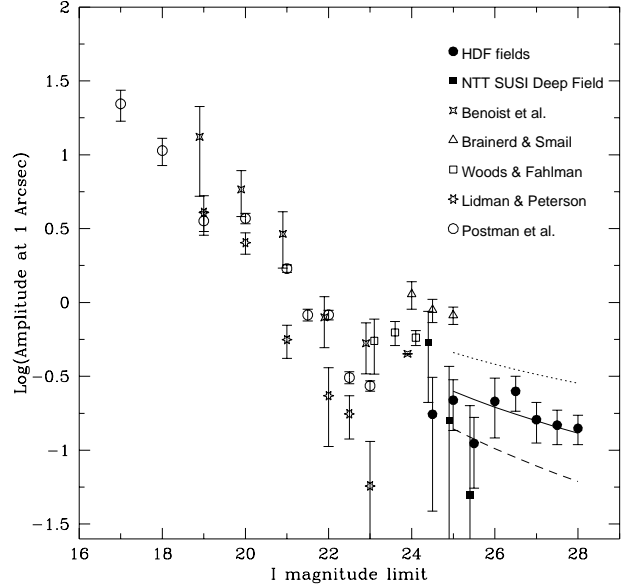,width=8.5cm}
\caption{The amplitude of $w(\theta)$ at 1 arcsec as a function of
the limiting I band magnitude for a number of ground based surveys and
the measurements from the Hubble Deep fields presented in this
paper. The superimposed curves are again the models with a
present clustering length of $r_0=4.0 h^{-1}$ Mpc and assuming no
evolution of clustering (dotted line), linear evolution (solid line)
as well as non-linear evolution of clustering (dashed line). }
\label{compare_groundI}
\end{figure}

\begin{itemize}

\item The clustering amplitudes measured in the HDF-N field have 
been confirmed with the independent sample from the Southern HDF. This 
rules out that the low amplitudes measured in the HDF-N field are due 
to some cosmological fluctuations. The HDF-N field seems to provide a 
fair sample of high-redshift galaxies. 
\item The clustering in  the HDF flanking fields is consistent with 
the one measured from ground based samples at the bright end of the
magnitude range, and with the ones measured in the HDFs at the faint 
end. This provides additional support to the view that the HDF fields 
represent indeed a fair view of the high-redshift universe. 
\item The measured correlation amplitudes in the ESO-NTT galaxy sample
in magnitude bins which overlap those of the HDF sample are consistent
with the HDF measurements. This observation verifies that correlations 
measured with WFPC2 are not due to some instrumental effects.
\item The results plotted in Fig.~\ref{compare_ground} show a 
continuous decline of the amplitude of the correlation function in R.
We do not confirm the Brainerd \& Smail (1998) detection of a flattening 
in the I band at I=22--23. 
\item There is some indication of a flattening of the
correlation amplitudes at the faintest levels ($R\approx27$ or
$I\approx26$). Using models for the redshift distribution, VFC interpreted
this flattening as evidence for linear evolution of the clustering of a
galaxy population which at present has a correlation length of about
4$h^{-1}$Mpc. The data presented in this paper are still in good
agreement with this interpretation.
\item Interpretation of Fig. 6 and 7 is difficult given the
significant disagreement between data points from different studies.
This disagreement could be caused by a change of $\delta$ as
a function of magnitude and scale.
To reach a final understanding of the two-point correlation function
data compiled in this work, it therefore appears to be essential first
to understand the detailed behavior of the {\it shape} of the
correlation function.

\end{itemize}


\section*{Acknowledgments}
We wish to thank C. Benoist and B. Thomsen for helpful discussions and
C. Benoist for giving
giving us access to some of his results from the EIS data prior to
publication. We thank S. D'Odorico for making available the catalog
from the NTT Deep Field. We thank the referee, T. Brainerd, 
for comments which significantly improved the
manuscript. This work made use of the ESO/ST-ECF Science Archive 
Facility.

\end{document}